\documentclass[conference]{IEEEtran}
\IEEEoverridecommandlockouts

\usepackage{cite}
\usepackage{amsmath,amssymb,amsfonts}
\usepackage{algorithmic}
\usepackage{graphicx}
\usepackage{textcomp}
\usepackage[table,x11names,dvipsnames,table]{xcolor}
\def\BibTeX{{\rm B\kern-.05em{\sc i\kern-.025em b}\kern-.08em
    T\kern-.1667em\lower.7ex\hbox{E}\kern-.125emX}}

\begin{document}

\title{Edge-Aligned Initialization of Kernels for Steered Mixture of Experts}

\author{\IEEEauthorblockN{Martin Determann and Elvira Fleig}
\IEEEauthorblockA{\textit{Communication Systems Group} \\
\textit{Technische Universität Berlin}\\
Berlin, Germany}\\Contact: m.determann@campus.tu-berlin.de}

\maketitle
\begin{abstract}
Steered Mixture of Experts (SMoE) has recently emerged as a powerful framework for spatial-domain image modeling, enabling high-fidelity image representation using a remarkably small number of parameters. Its ability to steer kernel-based experts toward structural image features has led to successful applications in image compression, denoising, super-resolution, and light field processing. However, practical adoption is hindered by the reliance on gradient-based optimization to estimate model parameters on a per-image basis—a process that is computationally intensive and difficult to scale.

Initialization strategies for SMoE are an essential component that directly affects convergence and reconstruction quality. In this paper, we propose a novel, edge-based initialization scheme that achieves good reconstruction qualities while reducing the need for stochastic optimization significantly. Through a method that leverages Canny edge detection to extract a sparse set of image contours, kernel positions and orientations are deterministically inferred. A separate approach enabls the direct estimation of initial expert coefficients. This initialization reduces both memory consumption and computational cost.
\end{abstract}

\section{Introduction}

The JPEG standard~\cite{Wallace_1991} remains the dominant method for natural image compression, leveraging block-based frequency-domain transforms. While efficient and widely supported, JPEG introduces visible artifacts and struggles to preserve structural detail under strong compression. As demands increase for high-resolution, low-latency, and content-adaptive representations, alternatives to transform coding are gaining traction~\cite{tok2018}.

One promising direction is the Steered Mixture of Experts (SMoE) framework~\cite{Verhack2016}, which models images as spatially localized Gaussian kernels steered along salient features. SMoE enables compact, interpretable representations and has shown promise in compression, denoising, and super-resolution. However, its practical deployment is hindered by the computational burden of parameter estimation. Existing methods rely on stochastic optimization to estimate kernel parameters—position, scale, orientation, and weights—resulting in long runtimes, high memory usage, and dependence on GPU acceleration.

Recent improvements, such as data-driven estimation~\cite{fleig2023, fleig20232}, structured initialization~\cite{li2024}, and regularization~\cite{Bochinski2018}, have mitigated some of these issues but still depend on iterative optimization and often impose rigid constraints on memory layout or tile sizing.

The most closely related method to ours~\cite{li2024} introduces an MDBSCAN-based segmentation followed by tile-wise optimization with random kernel placement. While adaptive, this approach is highly stochastic and computationally intensive.

In contrast, we propose a direct, deterministic initialization method that bypasses gradient descent during setup. Using Canny edge detection~\cite{Canny1986} and structural parsing, we place kernels orthogonally to dominant contours, apply simple filtering to avoid redundancy, and compute coefficients via closed-form estimation. This procedure executes in significantly less time, and enables flexible memory allocation and parallelism as it inherently supports arbitrary tile sizes. Although refinement via gradient descent is still necessary, we achieve reconstruction quality approaching state-of-the-art methods~\cite{li2024,li20242} at a fraction of the computational cost.
\section{Theoretical Background}

The Steered Mixture of Experts (SMoE) model is a spatial-domain regression framework for image representation and compression, designed to capture both smooth intensity variations and sharp transitions without introducing typical transform-domain artifacts~\cite{tok2018}. The model defines a function $L: \underline{x} \mapsto \mathbb{R}$, where $\underline{x} = (x, y)$ denotes pixel coordinates and $L(x,y)$ the corresponding grayscale intensity. The regression is expressed as

\begin{equation}
\label{ch1:eqn:smoe}
L(x,y) = \sum_{i=1}^K m_i \cdot w_i(x,y), \quad w_i(x,y) = \frac{\mathcal{K}_i(x,y)}{\sum_{j=1}^K \mathcal{K}_j(x,y)},
\end{equation}

with each kernel given by

\begin{equation}
\mathcal{K}_i(x,y) = \frac{1}{\alpha} \exp\left(-(\underline{x}-\underline{\mu}_i)^\top \boldsymbol{\Sigma}_i (\underline{x}-\underline{\mu}_i)\right),
\end{equation}

where $\underline{\mu}_i$ denotes the center, $\boldsymbol{\Sigma}_i$ the steering matrix encoding orientation and scale, and $\alpha$ a normalization constant. The weights $w_i(x,y)$ form a softmax function over the kernels and define a spatial gating mechanism. Each expert $m_i$ represents the amplitude or slope of a local hyperplane. To ensure positive semidefiniteness while maintaining computational efficiency, $\boldsymbol{\Sigma}_i$ is parameterized as $\boldsymbol{\Sigma}_i = A_iA_i^\top$ for $A_i \in \mathbb{R}^{2\times2}$, equivalent to the Cholesky decomposition. Parameters are optimized by minimizing the mean squared error between the reconstructed image and the ground truth:

\begin{equation}
\label{eqn:smoe:optim}
\min_{\{m_i, \underline{\mu}_i, \boldsymbol{\Sigma}_i\}} \; \mathcal{L}_{\text{MSE}}(L(x, y), I(x, y)),
\end{equation}

where $I(x, y)$ is the original image. Initialization can be achieved via uniform kernel placement or heuristics that allocate more components to structurally complex regions. The SMoE model has been shown to outperform standard codecs including JPEG, JPEG2000, and HEVC-Intra in terms of PSNR and SSIM at comparable bitrates~\cite{jongbloed2019}, owing to its ability to align model components with image structure.
\section{Edge-Aligned Initialization of SMoE}
\label{sec:theory}

We propose a lightweight, edge-informed, deterministic initialization strategy for Steered Mixture of Experts (SMoE) models that minimizes computational overhead and memory usage. By exploiting structural sparsity in natural images, extracted via edge detection and directional segmentation, this approach facilitates efficient model setup.

\begin{figure}[hbt!]
\centering
\includegraphics[width=\linewidth]{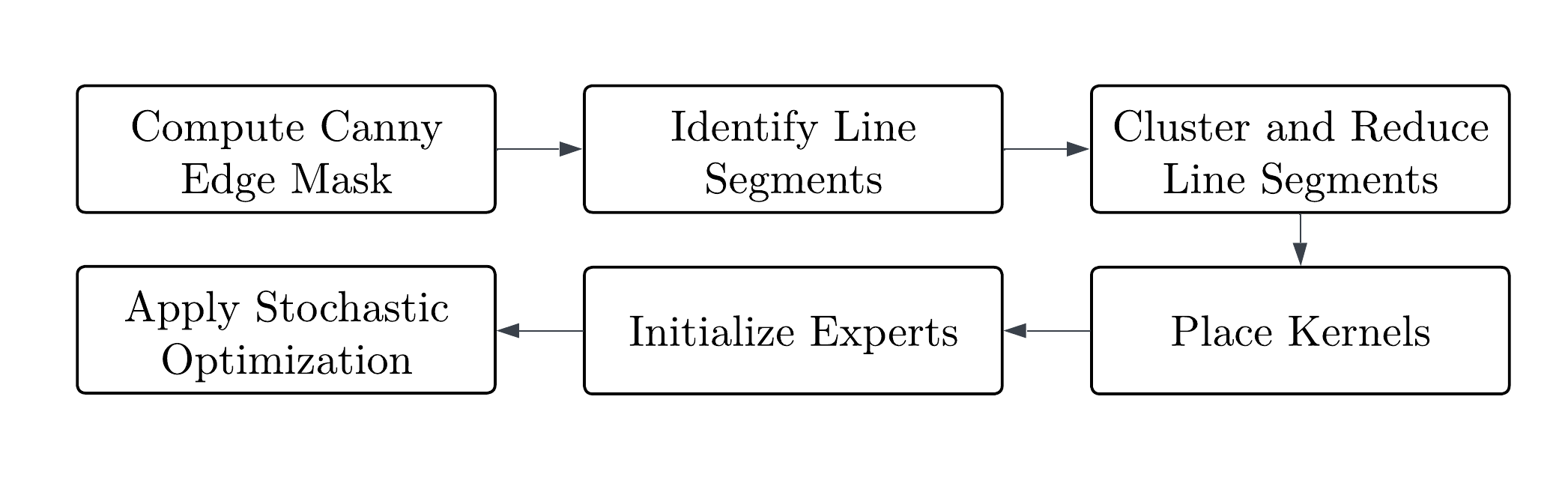}
\caption{Overview of the proposed SMoE initialization pipeline. Structural features are extracted via Canny edge detection and converted to a compact line-segment representation. This guides kernel placement and enables direct, gradient-free initialization of expert coefficients. A final refinement via gradient descent produces high-quality reconstructions.}
\label{fig:process_pipeline}
\end{figure}

As shown in Fig.~\ref{fig:process_pipeline}, the pipeline begins with Canny edge detection to identify prominent contours. These are converted into a line-segment representation aligned with the SMoE kernel structure. A reduction step regulates kernel count to enable adjustable compression rates. Kernels are then placed orthogonally to these segments. A brief iterative phase subsequently refines the expert values. Following this initialization, any tiled stochastic optimization may proceed efficiently.

\subsection{Canny Filtering and Line Segment Identification}
\begin{figure}[hbt!]
\centering
\includegraphics[width=0.25\linewidth,height=0.25\linewidth]{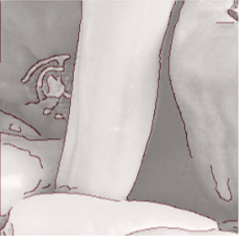}%
\includegraphics[width=0.25\linewidth,height=0.25\linewidth]{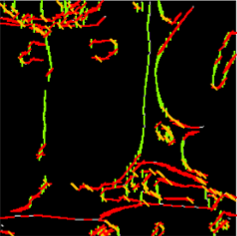}%
\includegraphics[width=0.25\linewidth,height=0.25\linewidth]{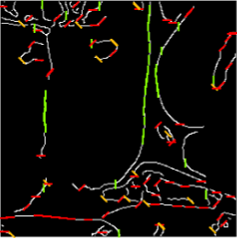}%
\includegraphics[width=0.25\linewidth,height=0.25\linewidth]{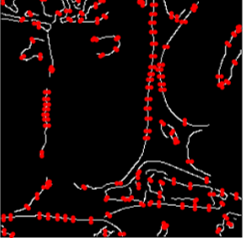}
(a)\hspace{0.21\linewidth}(b)\hspace{0.18\linewidth}(c)\hspace{0.21\linewidth}(d)
\caption{(a) Canny edge mask. (b) Initial line segments. (c) Segments after reduction. (d) Final kernel centers.}
\label{smoe:fig:line_finder_ex}
\end{figure}

Given a normalized grayscale image , the Canny detector \cite{Canny1986} extracts a binary edge mask, which provides an initial structural representation of the image. An example is shown in Fig.~\ref{smoe:fig:line_finder_ex}.
We introduce a directional line segmentation method applies the Canny edge mask \( E_b \) to derive a sparse set of oriented segments characterized by center and direction. In order to leverage the Canny edge mask \( E_b \) for a direct and optimal placement of Kernel means, it is necessary to extract a more geometrically sound representation of the mask. This representation must describe the mask in terms of structural components that are compatible with the SMoE model. To this end, we redefine the mask in terms of line segments. The line segments are extracted by scanning \( E_b \) along canonical directions \((0,1), (1,0), (1,1), (1,-1)\), corresponding to angles \( \theta \in \{0^\circ, 90^\circ, 45^\circ, -45^\circ\} \). A segment is initialized at a binary pixel \( E_b(x, y) = 1 \) with \( E_b(x - d_x, y - d_y) = 0 \), and extended to form a maximal connected sequence
\begin{equation}
\begin{aligned} 
S &= \{(x_k, y_k)\}_{k=1}^n\\
&s.t.\quad
    (x_k, y_k) = (x + k d_x, y + k d_y)\\
    &and\quad E_b(y_k, x_k) = 1.
\end{aligned}
\end{equation}
Segments with \( n \geq 2 \) are retained, and their geometric centers computed as:
\(
    \underline{\mu} = \frac{1}{2}[x_1 + x_n,y_1 + y_n]^\top.
\)
The resulting set of candidate kernel locations is given by:
\begin{equation}
    \mathcal{P} = \{ (\mu_i^x, \mu_i^y, \theta_i) \}_{i=1}^N.
\end{equation}

\subsection{Line Segment Reduction and Clustering}

Prior initialization methods determine the number of kernels solely based on initialization parameters, such as the Canny detector's sigma and threshold. While these parameters influence kernel quantity, they often lead to variability and redundancy, causing inconsistent parameter counts and increased optimization time proportional to these counts. To address this, we apply clustering to constrain the number of parameters while preserving representation quality.

For each candidate segment \( p_i \in \mathcal{P} \), we define an importance score \( s_i \in \mathbb{R} \) that accounts for spatial proximity and angular diversity. Let the neighborhoods of segments with similar and dissimilar orientations be:
\begin{equation}
\begin{aligned}
    \mathcal{N}_i^{\mathrm{sim}} &= \{ p_j \in \mathcal{P} \setminus \{p_i\} : \theta_j = \theta_i \}, \\
    \mathcal{N}_i^{\mathrm{dis}} &= \{ p_j \in \mathcal{P} : \theta_j \neq \theta_i \}.
\end{aligned}
\end{equation}
Define the average Euclidean distances to the two nearest neighbors in each set as:
\begin{equation}
\begin{aligned}
    d_i^{\mathrm{sim}} &= \text{mean}\left( \text{Top-2} \{ \|p_i - p_j\|_2 : p_j \in \mathcal{N}_i^{\mathrm{sim}} \} \right), \\
    d_i^{\mathrm{dis}} &= \text{mean}\left( \text{Top-2} \{ \|p_i - p_j\|_2 : p_j \in \mathcal{N}_i^{\mathrm{dis}} \} \right).
\end{aligned}
\end{equation}
The importance score is computed as
\begin{equation}
    s_i = (1 - \lambda) d_i^{\mathrm{sim}} + \lambda d_i^{\mathrm{dis}},
    \label{eq:importance}
\end{equation}
where \( \lambda \in [0,1] \) is a weighting hyperparameter, typically set to \(0.1\).

Segments with low scores are clustered using DBSCAN with spatial threshold \(\epsilon\). Each cluster \(\mathcal{C}_k\) is represented by the mean position and modal orientation:
\begin{equation}
    c_k = \frac{1}{|\mathcal{C}_k|} \sum_{p_i \in \mathcal{C}_k} p_i, \quad
    \bar{\theta}_k = \mathrm{mode}(\{\theta_i : p_i \in \mathcal{C}_k\}).
\end{equation}
To preserve salient features, the top \( K \) highest-scoring unclustered segments are retained. The final initialization set
\[
    \mathcal{K} = \{ (\tilde{\mu}_i^x, \tilde{\mu}_i^y, \tilde{\theta}_i) \}_{i=1}^M
\]
contains at most \( M \leq \texttt{max\_pts} \) segments.

The clustering radius \(\epsilon\) is iteratively increased until approximately \(0.8 \times \texttt{max\_pts}\) segments are clustered, leaving \(K = 0.2 \times \texttt{max\_pts}\) unique, high-scoring segments that are spatially distant from clusters. These isolated segments typically correspond to small but significant structures (e.g., pupils).

\subsection{Kernel Placement}

Kernels are placed directly using the line segmentation. For each segment, a pair of Gaussian kernels is positioned orthogonally to the segment orientation, separated by a fixed distance \(\Delta \mu\) [px]. Fig.~\ref{smoe:fig:line_finder_ex} illustrates kernel placement and segment identification. The kernel steering matrix \(\boldsymbol{\Sigma}\) is initialized isotropically as \(\boldsymbol{\Sigma} = \frac{1}{2\Delta \mu^2} I_{2 \times 2}\), where \(I\) is the identity matrix. Fig.~\ref{smoe:fig:line_finder_ex} (c) and (d) display this process visually.

\subsection{Expert Initialization}

\begin{figure}[hbt!]
\centering
\includegraphics[width=0.6\linewidth]{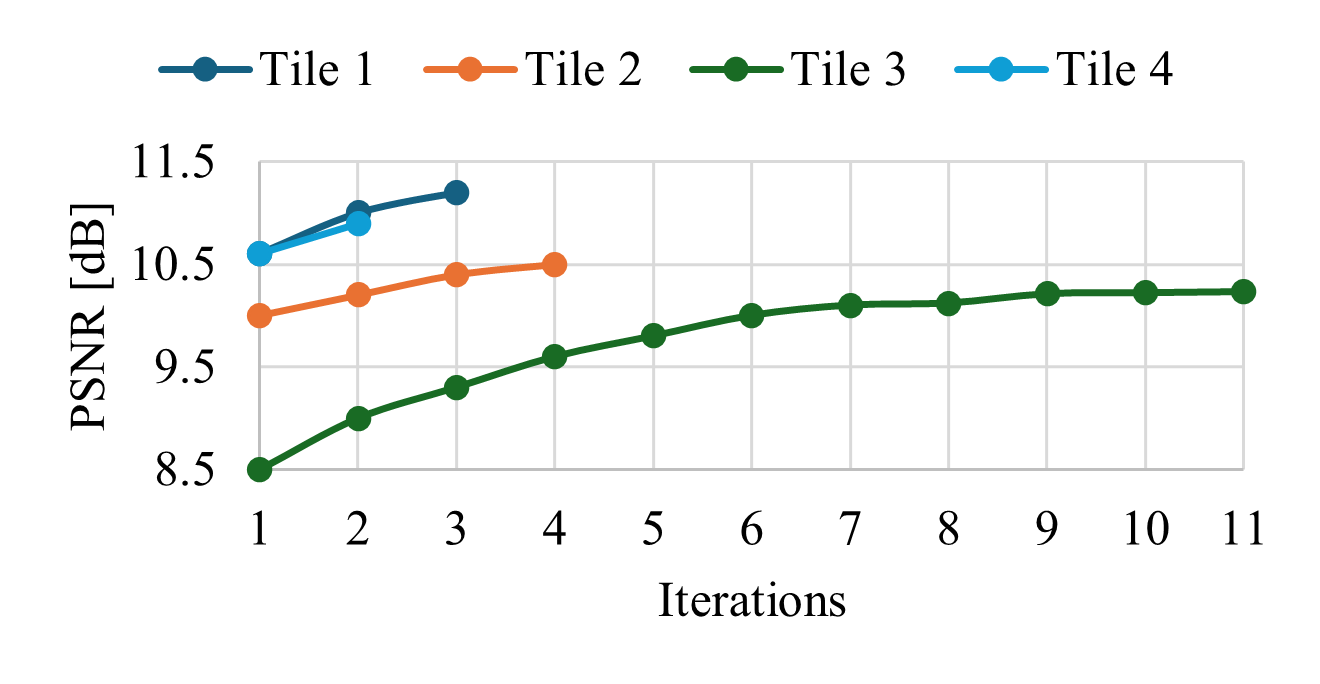}
\caption{
PSNR across expert initialization iterations.}
\label{fig:expert_init_pros}
\end{figure}

The expert parameters \( m \in [0,1]^d \), representing each kernel's contribution, are initialized by sampling the luminance image at kernel centers:
\begin{equation}
    m_n^{(0)} = L_n[\mu_n^x, \mu_n^y].
    \label{eqn:m_i_init}
\end{equation}
These values are refined iteratively via error minimization:
\begin{align}
    \hat{L}_n^{(i)} &= \mathrm{SMoE}(m_n^{(i)}, \mu_n^x, \mu_n^y, \boldsymbol{\Sigma}_n), \\
    m_n^{(i+1)} &= m_n^{(i)} + \eta \big(\hat{L}_n^{(i)}[\mu_n^x, \mu_n^y] - L_n[\mu_n^x, \mu_n^y]\big),
    \label{eqn:update_rule}
\end{align}
where \(\eta\) is a fixed step size (e.g., \(\eta = 0.1\)). Optimization proceeds for a fixed number of iterations or until convergence in mean squared error. Fig.~\ref{fig:expert_init_pros} demonstrates typical initialization results. This process ensures stability in subsequent model optimization.

\subsection{Stochastic Optimization}
\begin{figure}[hbt!]
\centering
\includegraphics[width=\linewidth]{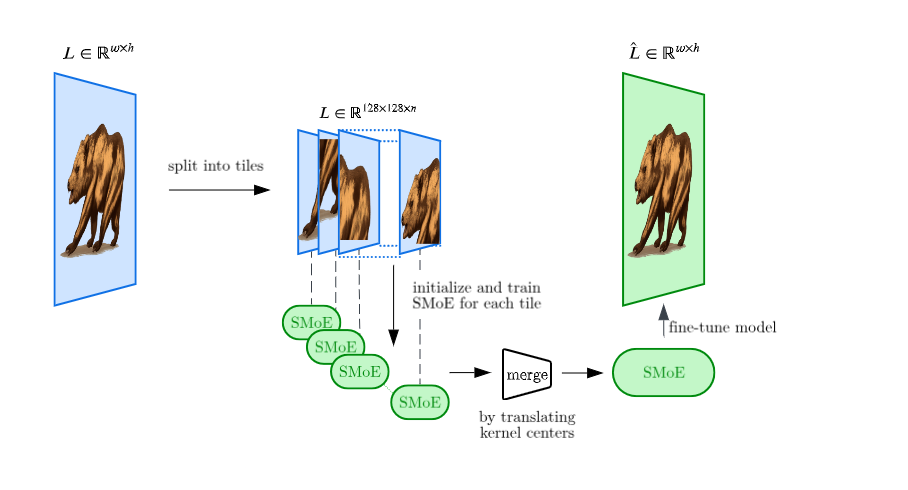}
\caption{Visualization of the procedure used to train the SMoE models for the test images. As with the S-SMoE and the AS-SMoE, the image is split into tiles, before training a SMoE to represent each tile~\cite{li2024,li20242}. Finally, the models are grouped by translating the kernel positions relative to the tile positions. The resulting model is then fine-tuned to represent the full image.}
\label{fig:experiments:model_splitting}
\end{figure}

The edge-informed initialization technique proposed in this work allows more flexibility in subsequent optimization, as the resulting models can readily be split into individual tiles to reduce memory consumption. To retain comparability with the state of the art S-SMoE~\cite{li2024} and AS-SMoE~\cite{li20242} initialization strategies, we employ a similar training strategy. We begin by splitting the original image into smaller tiles, before the proposed initialization technique is applied to create a SMoE Model for each tile. The models are trained with L2 regularization to minimize the normalization constants via $\left(\sum_i |\alpha_i|\right)^2$. The models are pruned during training by removing kernels as soon as their normalization constant $\alpha$ falls below a threshold. Next, the models are grouped into a single SMoE representing the full image, by translating the kernel centers. Fine-tuning without regularization is applied until convergence. The process is visualized in Fig.~\ref{fig:experiments:model_splitting}. 

\begin{figure*}[!t]
\centering
\includegraphics[width=\textwidth]{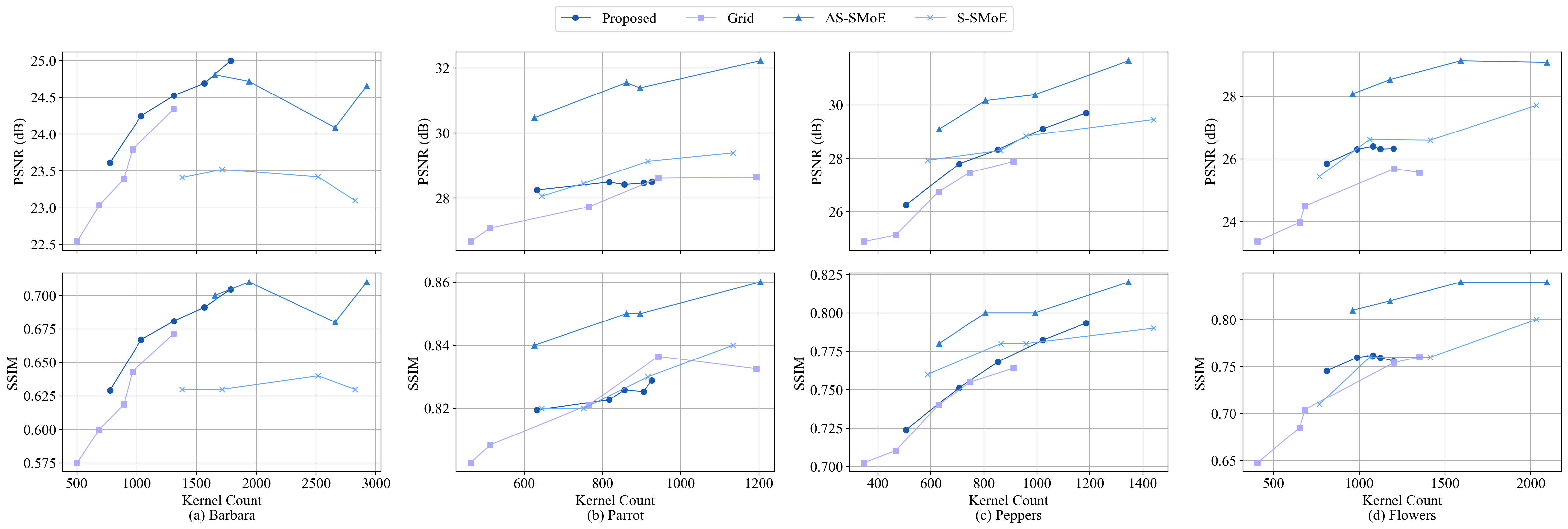}
\caption{PSNR vs. number of kernels for different initialization techniques. The proposed method  outperforms the grid baseline for the most part. The proposed method yields slightly poorer results than the S-SMoE and AS-SMoE techniques proposed in~\cite{li2024,li20242}.}
\label{fig:experiments:final_results_loss}
\end{figure*}

\begin{figure*}[!t]
\centering
\includegraphics[width=0.8\linewidth, trim={0 1cm 0 0}, clip]{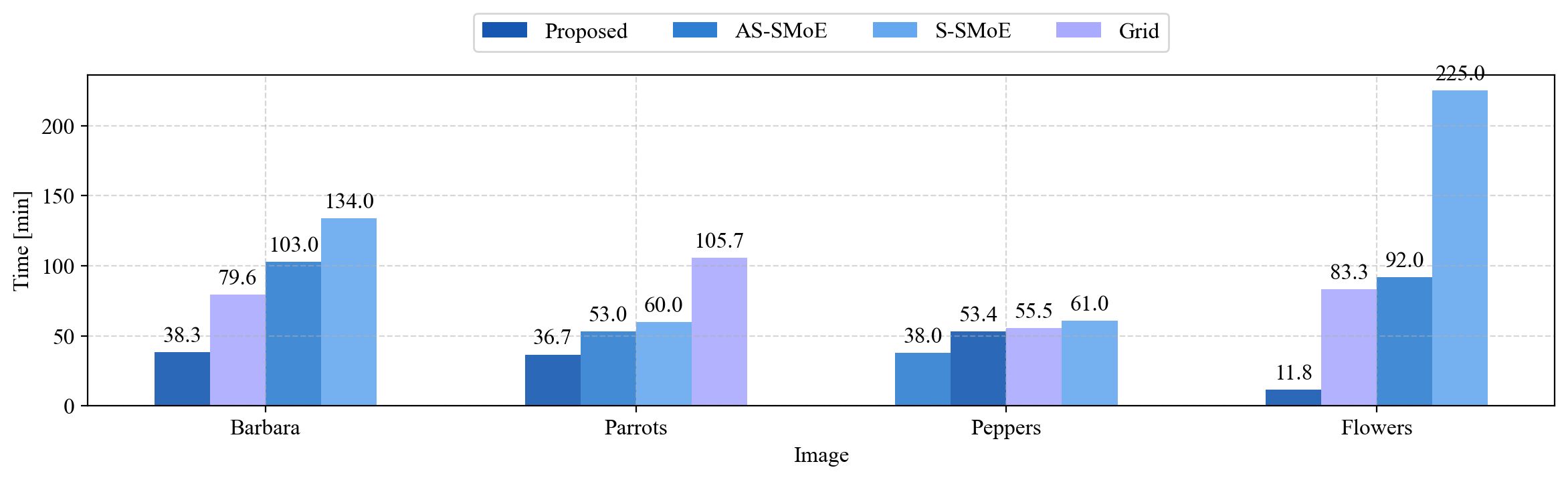}
\caption{Timing comparison between the four initialization techniques. The proposed spatial method outperforms the other methods in all images except for the peppers images, where the S-SMoE technique converged more quickly.}
\label{fig:experiments:final_results_time}
\end{figure*}
\section{Experiments}
\label{sec:experiments}

\begin{figure}[hbt!]
    \centering
    \includegraphics[width=\linewidth]{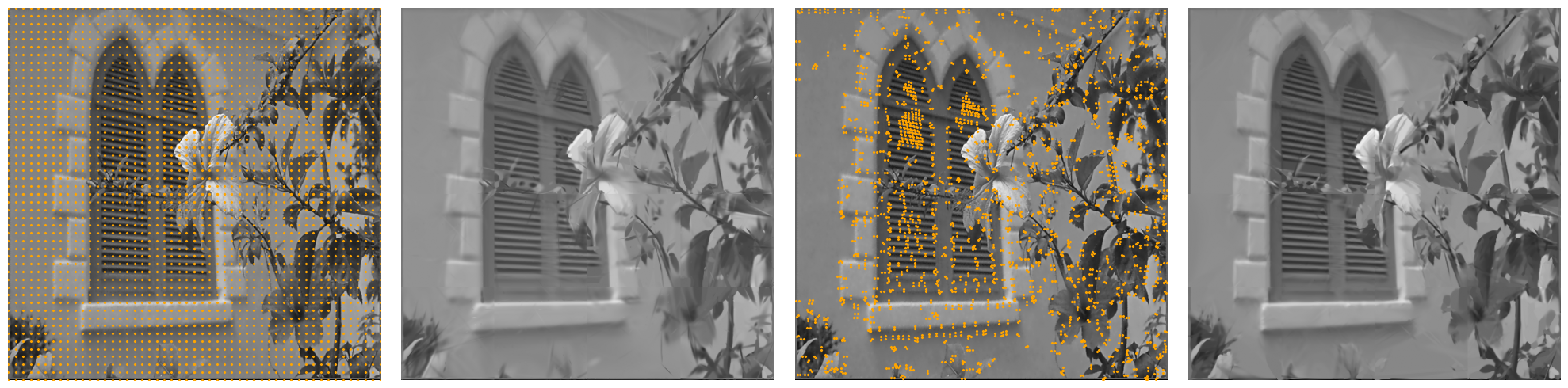}
    \caption{(a) Kernel centers from grid-based initialization. (b) SMoE reconstruction using grid initialization. (c) Kernel centers from proposed initialization. (d) SMoE reconstruction using proposed method.}
    \label{fig:experiments:dot_comparison}
\end{figure}

We evaluate the proposed initialization method within the SMoE framework described in Section~\ref{sec:theory}. Experiments are conducted on grayscale images Barbara, Flowers, and Parrots (\(768 \times 512\)) and Peppers (\(512 \times 512\)), normalized to \([0,1]\). Several initialization strategies are compared: (i) the proposed edge-aware approach using Canny edge priors, (ii) a uniform grid baseline distributing kernels evenly (Fig.~\ref{fig:experiments:dot_comparison}), (iii) the S-SMoE technique proposed in~\cite{li2024} and (iv) the AS-SMoE technique proposed in~\cite{li20242}.  

Qualitatively (Fig.~\ref{fig:experiments:dot_comparison}), the proposed method yields spatially adaptive, structured kernel placements that better preserve edge and texture details than the uniform grid on average. Quantitative results (Fig.~\ref{fig:experiments:final_results_loss}) show significant PSNR and SSIM improvements over the grid baseline. An increase in the number of kernels vs the baseline models can be attributed to the fact that the spatial placement of kernels results in their contribution to the reconstruction quality being greater than those in the baseline models. This reduces the impact of regularization and pruning during training and helps achieve better representations. The proposed method yields slightly lower PSNR and SSIM values than the state of the art S-SMoE and AS-SMoE initialization from~\cite{li2024,li20242} in terms of SSIM and PSNR. Timing results in Fig.~\ref{fig:experiments:final_results_time}a, however, show that the proposed method reduces total convergence time significantly. 
\break
\section{Conclusion}

We proposed a novel initialization method for SMoE image representations that reduces the computational overhead of parameter optimization by leveraging prior information towards an improved model initialization. Unlike prior approaches relying on gradient descent without explicitly incorporating spatial image structure, our method directly estimates optimal kernel positions and luminance values. By leveraging Canny edge detection combined with line segment extraction, we obtain a structured representation well-suited to the SMoE framework. A ranking scheme based on segment clustering assigns importance scores to kernel pairs, enabling consistent sparsity and adaptive kernel selection. This approach provides an interpretable, efficient initialization that accelerates convergence and improves representation quality. 

Future work will focus on extending the method to incorporate nonlinear segment modeling for enhanced kernel steering matrix initialization.
\clearpage
\bibliographystyle{plain} 
\bibliography{bib} 
\end{document}